\documentclass[floatfix,superscriptaddress,nofootinbib,aps,twocolumn]{revtex4}
\usepackage[dvips]{graphicx}
\usepackage{latexsym,amssymb,amsmath,epsfig,bm,times,psfrag}
\usepackage{color}
\usepackage[bookmarksnumbered,bookmarksopen,colorlinks,citecolor=red,linkcolor=blue]{hyperref}
\usepackage{epstopdf}
\renewcommand\b{\beta}

\renewcommand\r{\rho}

\renewcommand\o{\omega}
\newcommand\e{\epsilon}
\newcommand\g{\gamma}

\newcommand\m{\mu}
\newcommand\n{\nu}

\newcommand\p{\pi}

\newcommand\s{\sigma}
\newcommand\f{\phi}

\newcommand\ve{\varepsilon}

\renewcommand\L{\Lambda}

\newcommand{\fig}[1]{Fig.~\ref{#1}}

\newcommand{\sect}[1]{Sec.~\ref{#1}}

\newcommand\lb{\left(}
\newcommand\rb{\right)}
\newcommand\ls{\left[}
\newcommand\rs{\right]}

\newcommand\pt{\partial}

\newcommand{\bx}{{\vec x}}

\newcommand{\bp}{{\vec p}}

\newcommand{\bv}{{\vec v}}

\renewcommand{\part}{{\rm part}}

\renewcommand{\vec}{\boldsymbol}
\newcommand{\be}{\begin{equation}}
\newcommand{\ee}{\end{equation}}
\newcommand{\bear}{\begin{eqnarray}}
\newcommand{\eear}{\end{eqnarray}}
\newcommand{\ba}{\begin{array}}
\newcommand{\ea}{\end{array}}

\begin{document}

\title{Vorticity in low-energy heavy-ion collisions}

\author{Xian-Gai Deng}
\affiliation{Key Laboratory of Nuclear Physics and Ion-beam Application (MOE), Institute of Modern Physics, Fudan University, Shanghai 200433, China}
\author{Xu-Guang Huang\footnote{Corresponding author: huangxuguang@fudan.edu.cn}}
    \affiliation{Department of Physics and Center for Field Theory and Particle Physics, Fudan University, Shanghai, 200433, China}
    \affiliation{Key Laboratory of Nuclear Physics and Ion-beam Application (MOE), Institute of Modern Physics, Fudan University, Shanghai 200433, China}
    \author{Yu-Gang Ma}
    \affiliation{Key Laboratory of Nuclear Physics and Ion-beam Application (MOE), Institute of Modern Physics, Fudan University, Shanghai 200433, China}
    \affiliation{Shanghai Institute of Applied Physics, Chinese Academy of Sciences, Shanghai 201800, China}
     \author{Song Zhang}
    \affiliation{Key Laboratory of Nuclear Physics and Ion-beam Application (MOE), Institute of Modern Physics, Fudan University, Shanghai 200433, China}
    \date{\today}

\begin{abstract}
We study the kinematic and thermal vorticities in low-energy heavy-ion collisions by using the Ultra-relativistic Quantum Molecular Dynamics (UrQMD) model. We explore their time evolution and spatial distribution. We find that the initial vorticities have a non-monotonic dependence on the collision energy $\sqrt{s_{\rm NN}}$: as $\sqrt{s_{\rm NN}}$ grows the vorticities first increase steeply and then decrease with the turning point around $\sqrt{s_{\rm NN}}\sim 3-5$ GeV depending on the centrality.
\end{abstract}

\maketitle

\section{Introduction} \label{sec:intro}
Recently, the spin polarization of $\L$ and $\bar\L$ hyperons (``$\L$ polarization" hereafter) in Au + Au collisions was measured for the first time at RHIC~\cite{STAR:2017ckg} which confirmed the early idea discussed in Refs.~\cite{Liang:2004ph,Voloshin:2004ha,Gao:2007bc,Huang:2011ru}.
This measurement opened the door to a new realm of ``subatomic spintronics" in heavy-ion collisions. The underlying mechanism of the $\L$ polarization is the quantum mechanical coupling between spin and fluid vorticity~\cite{Becattini:2013fla,Becattini:2013vja,Fang:2016vpj,Florkowski:2018ahw} and, therefore, the measurement provides valuable information about the vorticity generated in heavy-ion collisions. The so-extracted vorticity averaged over the collision-energy range $\sqrt{s_{\rm NN}}=7.7 - 200$ GeV is of the order of $\o\sim 10^{22} s^{-1}$~\cite{STAR:2017ckg}, surpassing the vorticity observed in any other fluids and thus marking the creation of the ``most vortical fluid" in high-energy heavy-ion collisions.

The results reported in Ref.~\cite{STAR:2017ckg} and also in an earlier publication for $\sqrt{s_{\rm NN}}=200$ GeV only~\cite{Abelev:2007zk} are for the integrated $\L$ polarization (dubbed global polarization) at mid-rapidity which may reflect the global angular momentum of the colliding system. The subsequent measurements revealed more details of the local information of $\L$ polarization, including its dependence on the transverse momentum, azimuthal angle, and rapidity~\cite{Adam:2018ivw,Adam:2019srw}. These new measurements contain very nontrivial features that have attracted a lot of attention and discussions~\cite{Karpenko:2016jyx,Becattini:2016gvu,Becattini:2017gcx,Voloshin:2017kqp,Han:2017hdi,Sun:2017xhx,Li:2017slc,Shi:2017wpk,Sun:2018bjl,Wei:2018zfb,Xia:2018tes,Csernai:2018yok,Xu:2018fog,Hattori:2019lfp,Xia:2019fjf,Becattini:2019ntv,Florkowski:2019voj,Wu:2019eyi,Xie:2019jun,Guo:2019mgh,Guo:2019joy,Liu:2019krs,Xie:2019npz}. In particular, there exist unsolved discrepancies between the experiments and the theoretical calculations on the azimuthal-angle dependence of both the longitudinal and transverse polarization. In addition, the experiments also reported the measurement of the spin alignment of $\f$ and $K^{*0}$ mesons which show features that are also not fully understood~\cite{Liang:2004xn,Zhou:2019lun,Acharya:2019vpe}. These call for more detailed theoretical study of the vorticity and the spin-polarization phenomena in heavy-ion collisions.

One special feature of the measured global $\L$ polarization, $P_\L$, is its $\sqrt{s_{\rm NN}}$ dependence: the data shows that $P_\L$ increases when $\sqrt{s_{\rm NN}}$ decreases in the energy range $\sqrt{s_{\rm NN}}=7.7-200$ GeV, in opposite to the total angular momentum which decreases when $\sqrt{s_{\rm NN}}$ decreases. This trend extends to $\sqrt{s_{\rm NN}}=2.76$ and $5.02$ TeV according to the recent measurement by ALICE Collaboration~\cite{Acharya:2019ryw}. This has been understood from the fact that with larger $\sqrt{s_{\rm NN}}$ the fireball at mid-rapidity is closer to a Bjorken boost invariant fluid and allows smaller vorticity~\cite{Deng:2016gyh,Deng:2016yru}. But what if $\sqrt{s_{\rm NN}}$ decreases further into very low energy region with $\sqrt{s_{\rm NN}} < 7.7$ GeV? Apparently, at $\sqrt{s_{\rm NN}}\sim 2m_N$ ($m_N$ the mass of proton or neutron) the total angular momentum is nearly zero and therefore the vorticity (if it can be properly defined in such a situation; see \sect{sec:setup}) must also be small. This suggests that the vorticity may first grow and then fall down as $\sqrt{s_{\rm NN}}$ increases from $2m_N$ and the turning point may indicate the arising of a boost-invariant fluid at the mid-rapidity region. We note that, recently, the HADES Collaboration reported the measurement of $\L$ polarization at $\sqrt{s_{\rm NN}}=2.4$ GeV and found it consistent with zero albeit with a big uncertainty~\cite{HADES:2019}. Combining the measurements from STAR, ALICE, and HADES Collaborations, the $\L$ polarization indeed shows a non-monotonic dependence on $\sqrt{s_{\rm NN}}$ which first increases and then drops with $\sqrt{s_{\rm NN}}$ growing from very small to very large values.

The purpose of this paper is to study the fluid vorticity at low-energy heavy-ion collisions. This is complementary to the previous studies reported in Refs.~\cite{Jiang:2016woz,Deng:2016gyh,Wei:2018zfb,Becattini:2015ska} in which the fluid vorticity at high-energy heavy-ion collisions was studied. The present study may also provide certain theoretical backgrounds for the recent measurement of $\L$ polarization by HADES Collaboration~\cite{HADES:2019}. We note that the vorticities at NICA and FAIR energies were studied in Refs.~\cite{Teryaev:2015gxa,Xie:2016fjj,Ivanov:2017dff,Kolomeitsev:2018svb} which bear some overlap with the energy range that we will explore; but our main focus will be the energy dependence of the vorticities at low energies. The magnetic fields can also be generated in heavy-ion collisions~\cite{Huang:2015oca,Hattori:2016emy,Wang:2018ygc} which may lead to splitting between $\L$ and $\bar\L$ polarizations, which, however, will not be discussed in the present study. We will use natural unit $\hbar=k_B=c=1$.


\section{Numerical setup} \label{sec:setup}
We will study two different vorticities, the kinematic vorticity and the thermal vorticity. They are defined, in the tensor form, as
\begin{eqnarray}
\label{vor:kin}
\omega_{\m\n}=\frac{1}{2}\lb\pt_\n u_\m-\pt_\m u_\n\rb,\\
\label{vor:the}
\varpi_{\m\n}=\frac{1}{2}\lb\pt_\n \b_\m-\pt_\m \b_\n\rb,
\end{eqnarray}
where $u^\m=\g(1,\bv)$ is the fluid four-velocity with $\g=1/\sqrt{1-\bv^2}$ the Lorentz factor and $\b^\m=\b u^\m$ with $\b=1/T$ the inverse temperature. Note that the thermal vorticity is dimensionless. The corresponding vector-form vorticities are defined by
\begin{eqnarray}
\label{vor:kin2}
\omega^{\m}=-\frac{1}{2}\e^{\m\n\r\s}u_\n\o_{\r\s},\\
\label{vor:the2}
\varpi^{\m}=-\frac{1}{2}\e^{\m\n\r\s}u_\n\varpi_{\r\s}.
\end{eqnarray}
The kinematic vorticity $\o^\m$ is a natural covariant generalization of the usual vorticity $\vec\o=(1/2)\vec\nabla\times\bv$ which measures the local angular velocity of the fluid. The importance of the thermal vorticity relies on the fact that at global equilibrium it determines the spin polarization density of the fluid~\cite{Becattini:2013fla,Fang:2016vpj}.

In order to compute the vorticities, we need to first compute the fluid velocity and the temperature. In our simulation we will use the Ultra-relativistic Quantum Molecular Dynamics (UrQMD) model to obtain the position and momentum of each particle after the collision. The UrQMD model is a microscopic model extensively used in simulating the (ultra)relativistic heavy ion collisions; see Refs.~\cite{Bass:1998ca,Bleicher:1999xi,Petersen:2008kb} for detailed description of the UrQMD model. We have also checked the results for $\sqrt{s_{\rm NN}}\lesssim 2.5$ GeV by using the Isospin-dependent Quantum Molecular Dynamics (IQMD) model~\cite{Hartnack:1997ez,Xu:2016lue,Zhang:2017esm} and find no qualitative difference between the two models once the mean-field effects are properly included in the UrQMD model.


We then use these information to define the three-velocity by~\cite{Deng:2016gyh}
\begin{eqnarray}
\label{def:vel}
\bv(x)=\frac{\sum_{i=1}^N(\bp_i/E_i)\r(x,\bx_i)}{\sum_{i=1}^N \r(x,{\vec x}_i)},
\end{eqnarray}
where $\bp_i$ and $E_i$ are the momentum and energy of the $i$th particle located at $\bx_i(t)$, $N$ is the total particle number, and $\r(x,\bx_i)$ is a smearing function. We choose a Gaussian form for $\r$,
\begin{eqnarray}
\label{smear}
\r(x,\bx_i)=\frac{1}{(2\p\s^2)^{3/2}}\exp{\ls-\frac{(\bx-\bx_i(t))^2}{2\s^2}\rs},
\end{eqnarray}
where we choose the width parameter $\s$ to be $\s=1.48$ fm for baryons~\cite{Hartnack:1997ez} and $\s=0.98$ fm for mesons from a constituent quark number scaling for volume $\s_{ \p}=(2/3)^{1/3}\s_{\rm p,n}$. The energy density $\ve(x)$ is obtained similarly
\begin{eqnarray}
\label{def:ener}
\ve(x)=\sum_{i=1}^N E_i\r(x,\bx_i).
\end{eqnarray}

As we are considering the low-energy collisions in which the matter after the collisions may not reach the local equilibrium so that, in principle, the notation of temperature may not apply. Thus, we will use the ``temperature" $T(x)$ simply as a measure of the energy density, $\ve(x)$, via the relation $\ve=c T^4$; the concrete value of the prefactor $c$ is not essential for our purpose but we choose it to be $c=\p^2(16+10.5N_f)/30\approx 15.6$ with $N_f=3$ so that it can recover the $\ve-T$ relation for the thermalized quark-gluon matter in high-energy collisions.

\section{Numerical Results} \label{sec:result}
\begin{figure}
\centering
\includegraphics[width=1.0\columnwidth]{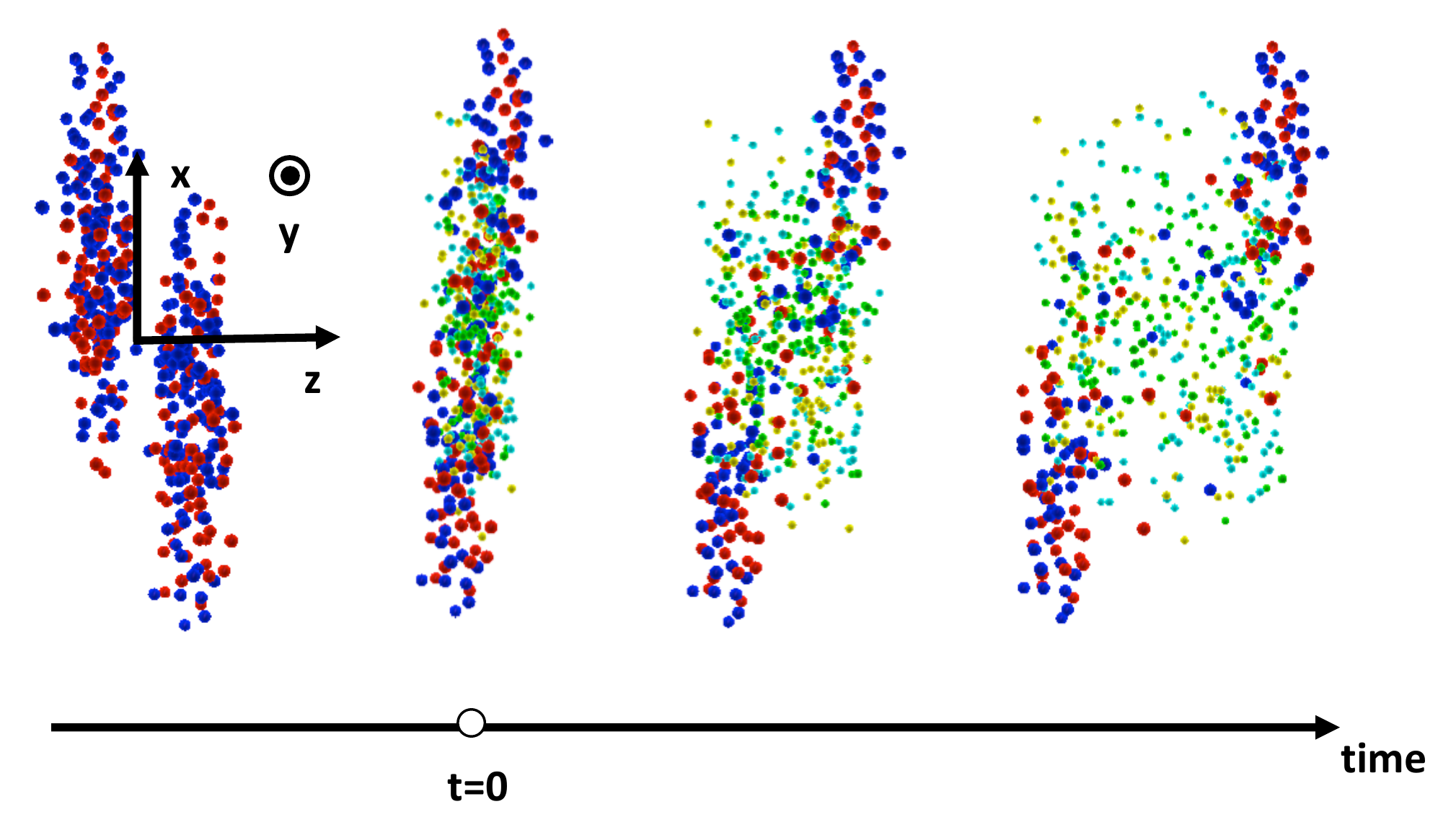}
\caption{Illustration of the time evolution of a Au + Au collision. The coordinate system is also shown.}
\label{fig:illu}
\end{figure}

We present our numerical results for both the kinematic and thermal vorticities for Au + Au collisions at $\sqrt{s_{\rm NN}}=1.9-50$ GeV. 
In \fig{fig:illu} we show an illustration of the evolution of the collision, the origin of the temporal axis is set to the moment when the number density of the particles is maximized in the beam direction.

\begin{figure}
\centering
\includegraphics[width=1.0\columnwidth]{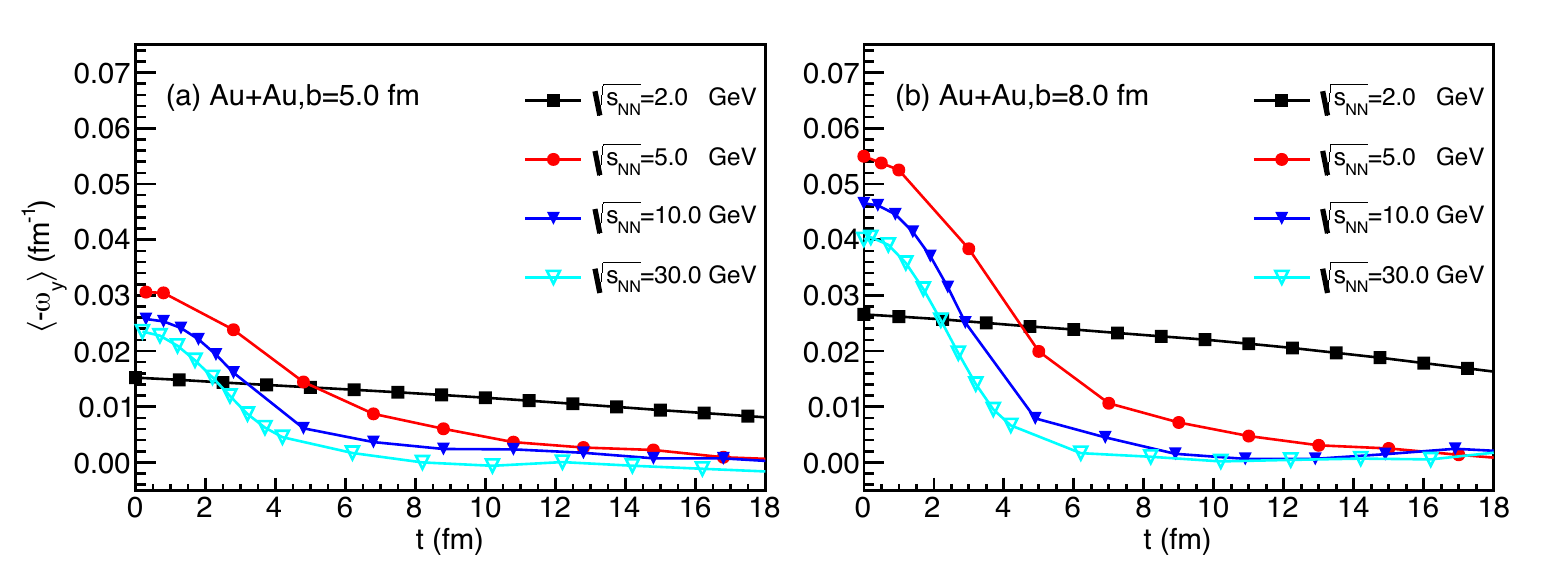}
\caption{Time evolution of the mid-rapidity kinematic vorticity at different energies and two different impact parameters in the UrQMD model.}
\label{fig:vort-time}
\end{figure}

At mid-rapidity, the vorticity at the center of the collision region (i.e., $\bx={\vec 0}$) is along the direction of the total angular momentum, i.e, the $-y$ direction in \fig{fig:illu}. The results for the kinematic vorticity for impact parameter $b=5, 8$, and $10$ fm are shown in \fig{fig:vort-time}. The average denoted by $\langle\cdots\rangle$ is over the overlapping region with weight $\ve$ and over 500 events~\cite{Deng:2016gyh}. The results show that the kinematic vorticity decays with time as a result of the system expansion with decay rate faster at higher energy as the system expands faster at higher energy.
An important feature seen in \fig{fig:vort-time} is that in the energy range $\sqrt{s_{\rm NN}}\lesssim 5$ GeV, the kinematic vorticity at $t=0$ (which will be called initial vorticity) grows with $\sqrt{s_{\rm NN}}$. This is very different from that for high-energy collisions where it is already known that the kinematic vorticity decreases with increasing $\sqrt{s_{\rm NN}}$~\cite{Jiang:2016woz,Deng:2016gyh}.

\begin{figure}
\centering
\includegraphics[width=0.82\columnwidth]{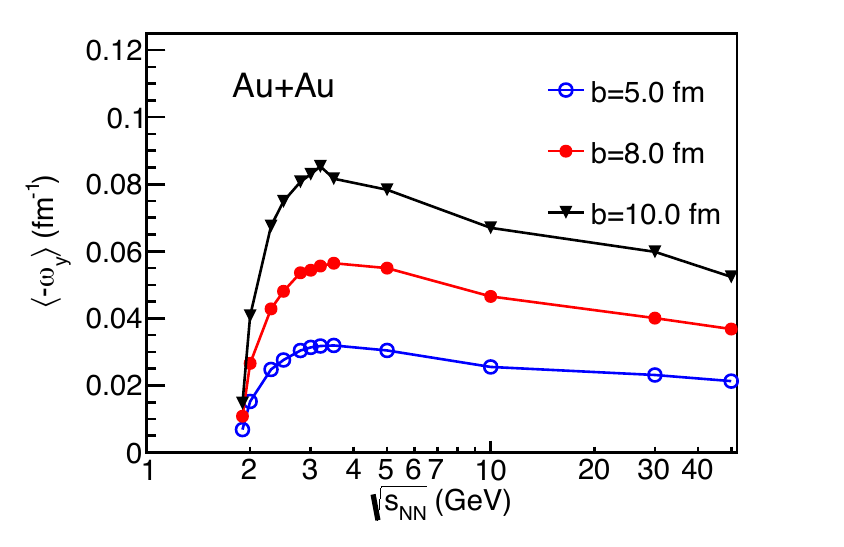}
\caption{Initial kinematic vorticity at mid rapidity as a function of the collision energy for impact parameters $b=5, 8$, and $10$ fm.}
\label{fig:vor-ene}
\end{figure}

This is more clearly seen in \fig{fig:vor-ene}: The initial kinematic vorticity versus $\sqrt{s_{\rm NN}}$ is non-monotonic. With $\sqrt{s_{\rm NN}}$ grows from $2m_N$, a finite angular momentum also grows which induces a finite kinematic vorticity increasing with $\sqrt{s_{\rm NN}}$; namely, most of the angular momentum is carried by the particles near the mid-rapidity region. When $\sqrt{s_{\rm NN}}$ is large enough (our computation suggests a turning point around $\sqrt{s_{\rm NN}}\sim 3-5$ GeV depending on centrality), the particles near the mid-rapidity are not effective angular-momentum carriers and most of the angular momenta are carried by the particles with large rapidity (but at large rapidity the angular momentum may not be necessarily manifested as fluid vorticity) and leaving the mid-rapidity region approximately boost invariant. With $\sqrt{s_{\rm NN}}$ growing to be very large, the mid-rapidity region respects a good Bjorken scaling structure which does not support the fluid vorticity. We note that in recent preliminary results reported by HADES Collaboration~\cite{HADES:2019}, the $\L$ polarization indeed appears to be very small at $\sqrt{s_{\rm NN}}=2.4$ GeV. Recalling that the global $\L$ polarization at $\sqrt{s_{\rm NN}}=7.7-200$ GeV measured by STAR Collaboration~\cite{STAR:2017ckg} and at $\sqrt{s_{\rm NN}}=2.76$ and $5.02$ TeV by ALICE Collaboration~\cite{Acharya:2019ryw} is decreasing with $\sqrt{s_{\rm NN}}$, our results combined with the previous studies in, e.g. Ref.~\cite{Deng:2016gyh}, are consistent with the current experimental data if we adopt the vorticity interpretation of the global $\L$ polarization.

\begin{figure}
\centering
\includegraphics[width=1.0\columnwidth]{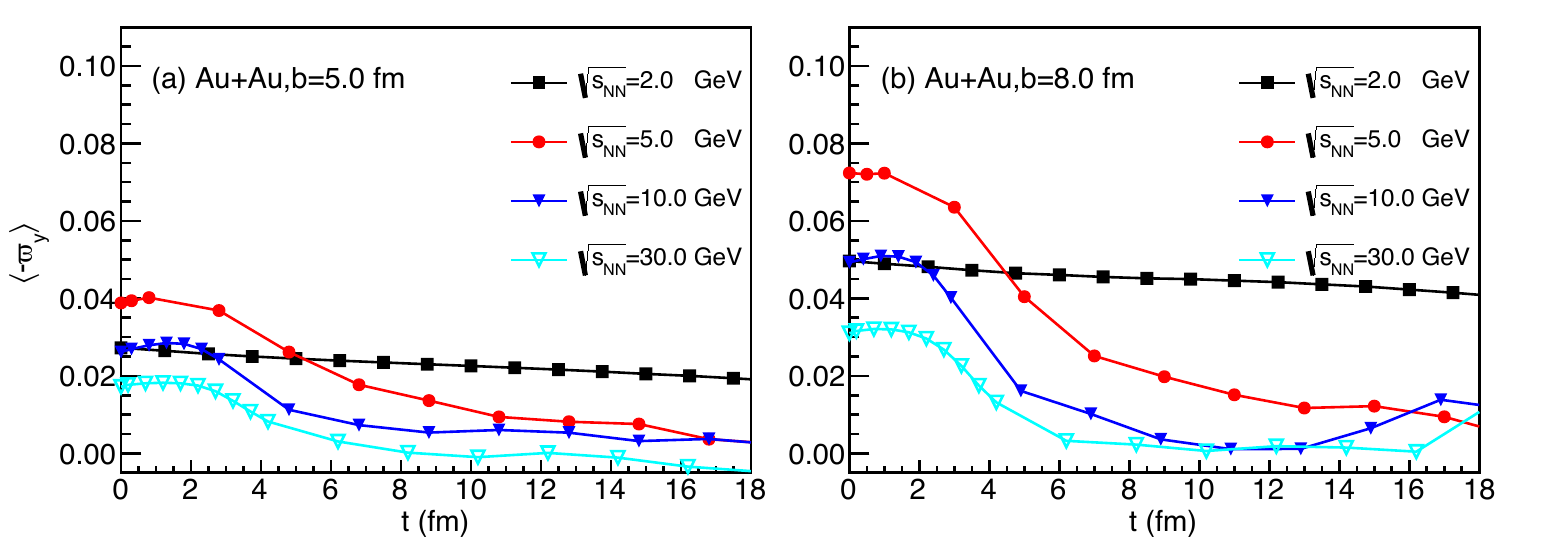}
\caption{Time evolution of the mid-rapidity thermal vorticity at different energies and impact parameters in the simulation with the UrQMD model.}
\label{fig:ther-time}
\end{figure}

We show the time evolution of the thermal vorticity in \fig{fig:ther-time} for two different centralities given by $b=5$ fm and $b=8$ fm. It exhibits similar time dependence comparing to \fig{fig:vort-time} for the kinematic vorticity. 
It was shown that if a fluid is at global equilibrium the thermal vorticity is responsible for determining the spin polarization density of the fluid~\cite{Becattini:2013fla,Fang:2016vpj,Hattori:2019lfp,Liu:2019}. In low-energy heavy-ion collisions, we must emphasize that the system may not reach thermal equilibrium and may not have a well-defined local temperature in the thermodynamic sense. Thus, the temperature and in turn the thermal vorticity shown in \fig{fig:ther-time} may not have the same physical meaning as that given in a system at equilibrium. So in this situation we do not expect that the thermal vorticity we show here can determine the spin polarization. However, it could still be regarded as the low-collision-energy counterpart of the thermal vorticity defined at high collision energy and thus can give some hint about the spin polarization at low collision energies.

\begin{figure}
\centering
\includegraphics[width=0.8\columnwidth]{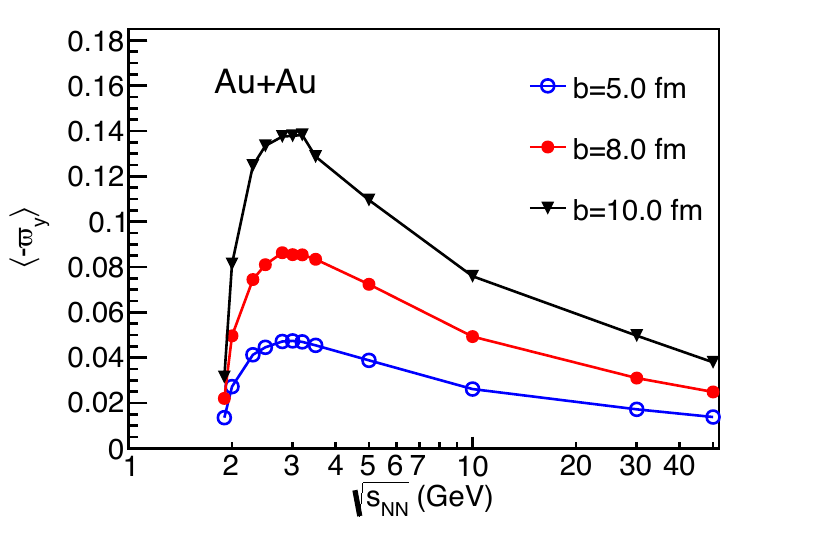}
\caption{Initial thermal vorticity at mid rapidity as a function of the collision energy for impact parameters $b=5, 8$, and $10$ fm.}
\label{fig:ther-ene}
\end{figure}

In parallel with \fig{fig:vor-ene}, we show the energy dependence of the thermal vorticity at mid-rapidity for Au + Au collisions in \fig{fig:ther-ene} which also exhibits non-monotonic feature. We here note that the energy dependence of the thermal vorticity at low-energy range was also calculated recently by using the three-fluid dynamics (3FD) model~\cite{Ivanov:2019ern}. They adopted a different definition for the origin of the time axis so that our vorticity at $t=0$ roughly corresponds theirs at the peak value; in this sense, their results are qualitatively consistent with ours. We note that although the initial thermal vorticity is non-monotonic, the thermal vorticity at late time (e.g., at $t=14$ fm) is roughly a decreasing function of $\sqrt{s_{\rm NN}}$. Given that the mean freeze-out times of $\L$ and $\bar{\L}$ hyperons may not be short even at low energy~\cite{Vitiuk:2019rfv}, this means that the $\L$ polarization may behave differently from that for the initial thermal vorticity as shown in \fig{fig:ther-ene}. However, because at $\sqrt{s_{\rm NN}}\sim 2m_N$ there is no angular momentum to polarize $\L$'s spin, we do expect a vanishing $\L$ polarization at $\sqrt{s_{\rm NN}}\sim 2m_N$ (as shown by HADES Collaboration~\cite{HADES:2019}) and thus a non-monotonic behavior of $\L$ polarization as a function of $\sqrt{s_{\rm NN}}$. The calculation of the actual $\L$ polarization at low energy deserves a future study.

\begin{figure}
\centering
\includegraphics[width=1.0\columnwidth]{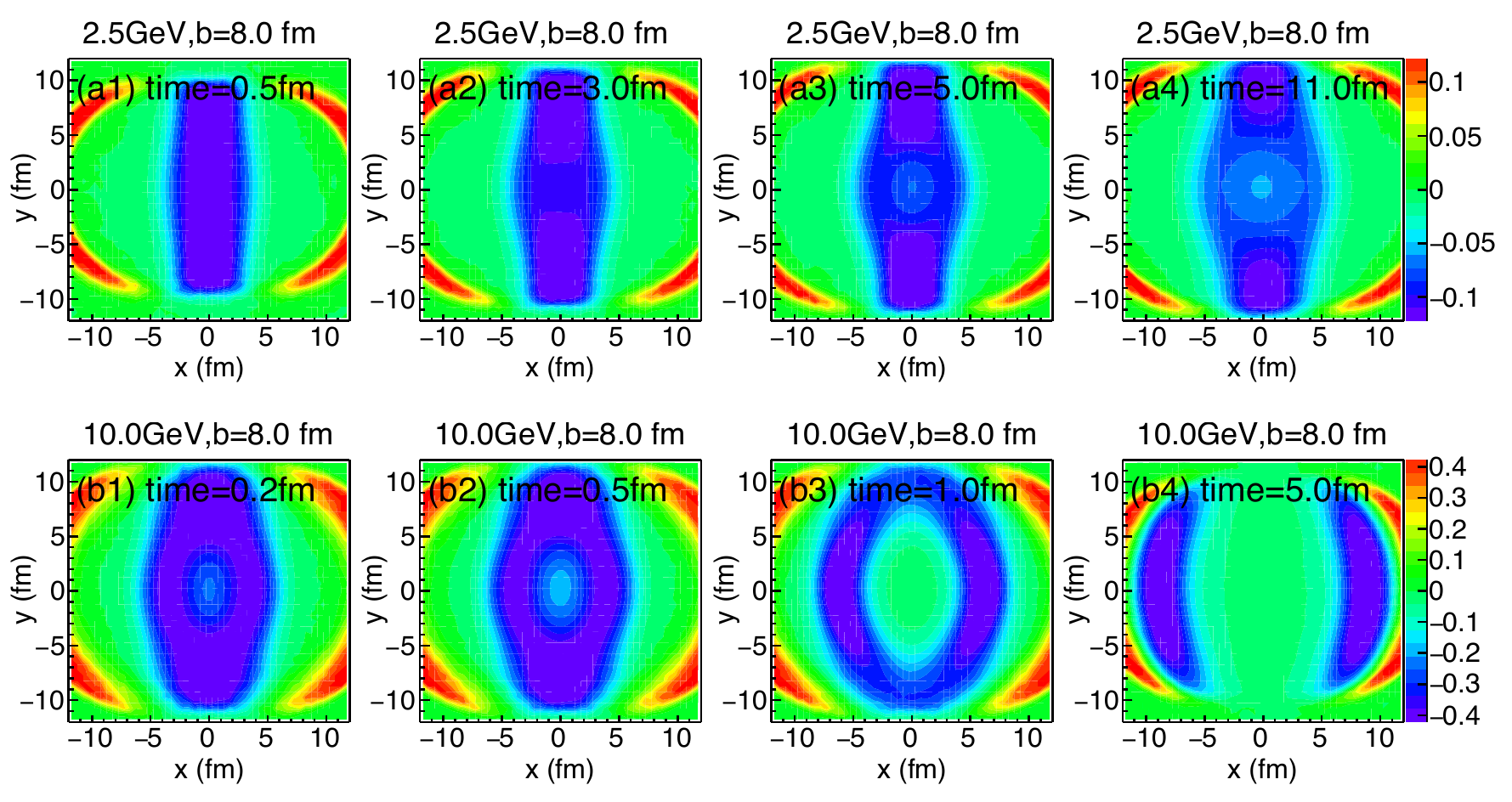}
\caption{The spatial distribution of the kinematic vorticity in the transverse plane for $\sqrt{s_{\rm NN}}=2.5$ GeV and $\sqrt{s_{\rm NN}}=10$ GeV for $b=8$ fm.}
\label{fig:spa-kin}
\end{figure}

\begin{figure}
\centering
\includegraphics[width=1.0\columnwidth]{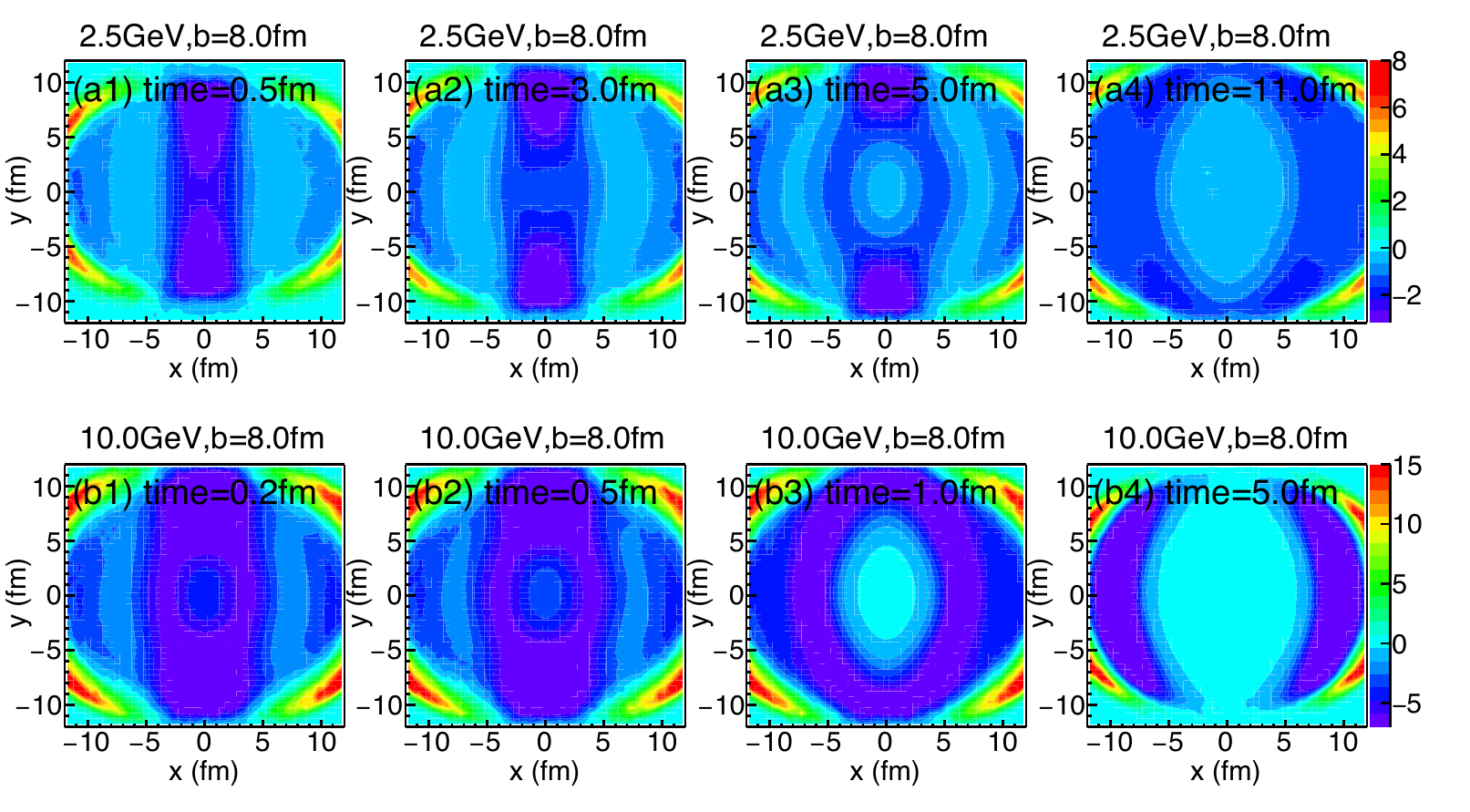}
\caption{The spatial distribution of the thermal vorticity in the transverse plane for $\sqrt{s_{\rm NN}}=2.5$ GeV and $\sqrt{s_{\rm NN}}=10$ GeV for $b=8$ fm.}
\label{fig:spa-the}
\end{figure}

Finally, we show the spatial distribution of the vorticities in the transverse plane, i.e. the $x$-$y$ plane, in \fig{fig:spa-kin} and \fig{fig:spa-the}. We can observe from \fig{fig:spa-kin} that the kinematic vorticity is roughly negative in the overlapping region consistent with the direction of the angular momentum. As the system expands, the vorticity at the center of the overlapping region becomes smaller and smaller; this is more clearly seen in the bottom panels for $\sqrt{s_{\rm NN}}=10$ GeV as the system expands faster than that of $\sqrt{s_{\rm NN}}=2.5$ GeV shown in the top panels. One may also notice that there are regions (near the periphery of the nuclei) with strong positive vorticity which is a corona effect due to the sharp density difference at the boundary. Very similar phenomena are also shown for the thermal vorticity in \fig{fig:spa-the}.

\section{Summary and Discussions} \label{sec:dis}
In summary, we have computed the kinematic and thermal vorticities in low-energy heavy-ion collisions in the energy range $\sqrt{s_{\rm NN}}=1.9-50$ GeV in the framework of the UrQMD model. The results show that both the initial kinematic and thermal vorticities first grow when $\sqrt{s_{\rm NN}}$ increase from $2m_N$ and then decrease for larger $\sqrt{s_{\rm NN}}$; the turning point is around $3-5$ GeV depending on the centrality. If we assume that the global $\L$ polarization is simply proportional to the initial thermal vorticity, our results suggest that the global $\L$ polarization versus $\sqrt{s_{\rm NN}}$ is not monotonic: it would first increase and then decrease as $\sqrt{s_{\rm NN}}$ grows. Such a feature is consistent with the recent measurements by HADES, STAR, and ALICE Collaborations. But we emphasize that the calculation of the actual $\L$ polarization needs more detailed study because that $\L$'s mean freeze-out time is not short and that at low energy we need non-equilibrium treatment of $\L$ polarization. The future experimental programs and facilities, such as the phase II of the beam energy scan program at RHIC, the FAIR at GSI, the NICA at Dubna, and the HIAF in China, could cover the energy range we have explored in this work and provide further results about the vorticity and spin polarization at low energies.

{\bf Acknowledgments.---}  This  work  is  partially  supported by  the  National  Natural  Science  Foundation  of  China under  Contract  Nos. 11890714, 11421505, 11535012, 11675041, and 11947217, the Strategic Priority Research Program of the CAS under Grant No. XDB34030200, and the China Postdoctoral Science Foundation Grant No.2019M661332.



\end{document}